\documentstyle[eqsecnum,aps]{revtex}
\begin{document}
\twocolumn
\title{
Resistivity as a function of temperature\\ for models with hot
spots on the Fermi surface.  }
\author{R. Hlubina\cite{byline} and T. M. Rice}
\address{Theoretische Physik,
Eidgen\"{o}ssische Technische Hochschule H\"onggerberg, CH-8093
Z\"urich, Switzerland}
\maketitle
\begin{abstract}
\noindent
We calculate the resistivity $\rho$ as a function of temperature $T$
for two models currently discussed in connection with high temperature
superconductivity: nearly antiferromagnetic Fermi liquids and models
with van Hove singularities on the Fermi surface. The resistivity is
calculated semiclassicaly by making use of a Boltzmann equation which
is formulated as a variational problem. For the model of nearly
antiferromagnetic Fermi liquids we construct a better variational
solution compared to the standard one and we find a new energy scale
for the crossover to the $\rho\propto T^2$ behavior at low
temperatures. This energy scale is finite even when the
spin-fluctuations are assumed to be critical. The effect of additional
impurity scattering is discussed.  For the model with van Hove
singularities a standard ansatz for the Boltzmann equation is
sufficient to show that although the quasiparticle lifetime is
anomalously short, the resistivity $\rho\propto T^2\ln(1/T)$.

\end{abstract}
\pacs{PACS numbers:}

\section{Introduction.}
One of the most challenging questions motivated by the anomalous
normal state of the high temperature superconductors (HTc's) is
to explain the linearity of the resistivity $\rho$ as a function
of temperature $T$ down to the superconducting transition
temperature $T_c$. In the case of the one-layer Bi-based
material, $T_c\approx 10 K$. This temperature is well below the
Debye temperature and, for a three-dimensional material, one
should expect that $\rho\propto T^5$ if the resistivity is
dominated by electron-phonon scattering.  Moreover,
electron-electron scattering (if treatable in perturbation
theory) is known to yield $\rho\propto T^2$ for temperatures
much less than the Fermi energy $E_F$.  Thus, conventional
theories do not explain the observed behavior of the resistivity
and one has to search for other mechanisms which could lead to a
higher scattering rate for the electrons.

Since superconductivity appears in the HTc's close to a
metal-insulator transition driven by strong correlations, it is
widely believed that some aspect of electron-electron
interactions should be responsible for the anomalous normal
state and, finally, also for the high superconducting transition
temperatures in the copper oxides.  Another feature specific to
the HTc's is their layered structure and the consequent
two-dimensional nature of the electronic states. The resulting
problem of interacting electrons in two dimensions has not been
solved yet and so far, only phenomenological theories attempting
to describe the low-energy behavior of such models are
available.  In the literature, two fundamentally different
routes for explaining the anomalies of the normal state of the
HTc's were followed. In one class of theories it is assumed that
in the HTc's, Landau's Fermi-liquid theory breaks down
completely and an exotic metallic state with some features of
one-dimensional solutions is realized \cite{Anderson,NL}.  In
the other class of theories, it is assumed that Landau's concept
of quasiparticles does apply. However, in order to explain the
deviations from the usual metallic behavior, anomalous
scattering mechanisms are assumed and treated in perturbation
theory.  In the present paper, we will study the temperature
dependence of the resistivity in two models of the latter type:
nearly antiferromagnetic Fermi liquids \cite{MP,Ueda} and models
with van Hove singularities \cite{Pattnaik,Markiewicz}.

In the theory of nearly antiferromagnetic Fermi liquids, it is assumed
that the effect of the strong local repulsion between electrons can be
described in the low-energy sector by coupling the electrons to an
overdamped low-lying paramagnon mode. The parameters of the
spin-fluctuation spectrum can be determined by fitting the magnetic
properties of the HTc's.  Standard weak-coupling calculations of the
resistivity in the model of nearly antiferromagnetic Fermi liquids
give $\rho\propto T$ for $T>T^\star$ where $T^\star$ measures the
deviation from the antiferromagnetic critical point.  More
sophisticated strong-coupling calculations of the resistivity support
the weak-coupling results in that there are only quantitative changes
to the latter. In Section 2, we elaborate the following observation:
since the spin-fluctuation spectrum is soft at the Brillouin-zone
boundary, only electrons in the vicinity of special points on the
Fermi line are strongly scattered. Along the rest of the Fermi line,
the lifetime of an electron is $1/\tau\propto T^2$ even for
$T>T^\star$.  Thus the contribution of the strongly scattering special
points to the resistivity is short-circuited by the remaining
electrons and the resistivity has the standard Fermi-liquid form
$\rho\propto T^2$ up to a new energy scale described in Section 2. A
similar idea was exploited by Fujimoto, Kohno, and Yamada
\cite{Fujimoto} in their discussion of the $T$-dependence of the
resistivity for models where parts of the Fermi surface exhibit
perfect nesting.

Another interesting proposal to explain the anomalies of the
HTc's is to assume that these are caused by the presence of van
Hove singularities at or close to the Fermi line. For instance,
the resistivity is enhanced according to these theories due to
the increase of phase space available in the scattering process.
Moreover, assuming a weak-coupling BCS formula for the
superconducting transition temperature, $T_c$ becomes enhanced
due to the large density of states at the Fermi level. It is
considered as a success of the van Hove scenario that the
anomalies of the normal state are strongest at that doping where
$T_c$ is maximal.  In Section 3 we show, however, that although
the single-particle lifetime is anomalous, the resistivity is
consistent with the standard result of Landau's Fermi-liquid
theory with a logarithmic correction, $\rho\propto T^2\ln(1/T)$.

\section{Nearly antiferromagnetic Fermi liquids.}
Following Monthoux and Pines \cite{MP}, we consider electrons
moving in a square lattice of Wannier orbitals with a simple
tight-binding spectrum
\begin{equation}
\varepsilon_{\bf k}=-2t\left[\cos(k_xa)+\cos(k_ya)\right]
+4t^\prime\cos(k_xa)\cos(k_ya),
\label{eq:MPdispersion}
\end{equation}
where $a$ is the lattice constant. $t=0.25 eV$ and
$t^\prime=0.45 t$ are nearest- and next-nearest-neighbor
hoppings, respectively.  The electrons are coupled to the
spin-fluctuation operator ${\bf S}$. The Hamiltonian reads
\begin{equation}
H=\sum_{{\bf k},\sigma}\varepsilon_{\bf k} c^\dagger_{{\bf
k},\sigma} c_{{\bf k},\sigma}+{{\bar g}\over 2}\sum_{{\bf
k,q},\alpha,\beta} c^\dagger_{{\bf k+q},\alpha}c^\dagger_{{\bf
k},\beta}\sigma_{\alpha,
\beta}\cdot {\bf S_{-q}},
\end{equation}
where $\bar g$ is a coupling constant.  The spectrum of the
spin-fluctuations is $\chi_{i,j}({\bf q},\omega)=\delta_{i,j}
\chi({\bf q},\omega)$ with
\begin{equation}
\chi({\bf q},\omega)=
{A\over{\omega_{\bf q}-i\omega}},
\label{eq:fluctuationspectrum}
\end{equation}
where $A\approx 1$ (in what follows we take $A=1$), $\omega_{\bf
q}=T^\star+\alpha T+\omega_D\psi_{\bf q}$ and $\psi_{\bf
q}=2+\cos(q_xa)+\cos(q_ya)$.  $T^\star$, $\alpha$, and
$\omega_D$ are temperature-independent parameters. Note that the
energy of the spin-fluctuations is minimal for ${\bf q}={\bf
Q}=(\pi/a,\pi/a)$ and $\omega_{\bf Q}= T^\star+\alpha T$. Thus
the parameter $T^\star$ measures the deviation from the critical
antiferromagnetic point.

\subsection{Quasiparticle lifetime.}

Before calculating the resistivity due to scattering on the
spin-fluctuations, let us analyze first the single-particle
lifetime.  In the second order of perturbation theory in $\bar
g$, the lifetime of an electron with momentum ${\bf k}$ at zero
temperature is
\begin{equation}
{1\over \tau_{\bf k}}=2g^2
\sum_{\bf k^\prime}\int_0^{\varepsilon_{\bf k}}d\omega\:
\Im \chi({\bf k}-{\bf k}^\prime,\omega)\:\delta(\varepsilon_{\bf k}-
\varepsilon_{\bf k^\prime}-\omega),
\label{eq:goldenrule1}
\end{equation}
where we have introduced $g^2=3{\bar g}^2/4$; $\Im\chi$ denotes
the imaginary part of $\chi$.  We can write $\int d^2k=\int
dk_\perp dk_\parallel$ where $k_\perp$ and $k_\parallel$ are the
directions perpendicular and parallel to the Fermi line. Since
$\int dk_\perp=\int d\varepsilon/ |{\bf v}|$ where ${\bf
v}=\nabla_{\bf k}\varepsilon$ is the group velocity we have
\begin{equation}
{1\over \tau_{\bf k}}=2\left({ga\over 2\pi}\right)^2\int_0^
{\varepsilon_{\bf k}} d\omega
\oint{dk^\prime\over v^\prime}\:
\Im\chi({\bf k-k^\prime},\omega),
\label{eq:goldenrule2}
\end{equation}
where we write $k^\prime$ instead of $k^\prime_\parallel$.  In
evaluating $\Im \chi({\bf k}-{\bf k}^\prime,\omega)$ we will
assume that both ${\bf k}$ and ${\bf k^\prime}$ lie on the Fermi
line, since corrections lead only to subleading terms in the
denominator of Eq.\ref{eq:fluctuationspectrum}. After
integrating over $\omega$ we have $${1\over \tau_{\bf
k}}=\left({ga\over {2\pi}}\right)^2\oint
{dk^\prime\over{v^\prime}}\ln\left( {{\varepsilon_{\bf
k}^2+\omega_{{\bf k}-{\bf k}^\prime}^2}\over{
\omega_{{\bf k}-{\bf k}^\prime}^2}}\right).$$
We assume that along the whole Fermi line the group velocity is
finite (the opposite case will be treated in Section 3).  It is
seen that $1/\tau_{\bf k}\propto\varepsilon_{\bf k}^2$ for
$\varepsilon_{\bf k}\ll T^\star$.  Thus $T^\star$ defines an
energy scale below which Landau's Fermi-liquid theory applies.
Let us investigate now the behavior of the lifetime for energies
$T^\star\ll\varepsilon_{\bf k}\ll\omega_D$.  Strong scattering
occurs only for those states ${\bf k}$ (hot spots) on the Fermi
line for which
\begin{equation}
\varepsilon_{\bf
k}=\varepsilon_{\bf k+Q}.
\label{eq:hotspot}
\end{equation}
For a wide range of fillings, there are only 8 points with
anomalous scattering (see Fig.1).  Note that for a commensurate
${\bf Q}=(\pi/a,\pi/a)$ the hot spots are by no means special
points of the Fermi line, since the group velocity in a point
${\bf k}$ satisfying Eq.\ref{eq:hotspot} is not parallel to
${\bf Q}$. The incommensurate case when ${\bf Q}$ is a locally
extremal vector connecting 2 points on the Fermi line (like the
$2k_F$-processes in continuum) will not be studied here.

Let $\delta {\bf k}$ be the distance between the projection of a
point ${\bf k}$ to the Fermi line and a nearby hot spot.  The
lifetime of an electron in the state ${\bf k}$ is
$${1\over\tau_{\bf k}}\propto{g^2\over E_F}\:
\sqrt{{\varepsilon_{\bf k}\over \omega_D}}\:
\min\left[1,\left({1\over \varphi}
\sqrt{{\varepsilon_{\bf k}\over \omega_D}}\:\right)^3\right],$$
where $\varphi=\delta k\:a$ and $E_F\sim v_F/a$; $v_F$ is the
Fermi velocity in the hot spot.  Thus, in agreement with our
expectations, the lifetime in the neighborhood of a hot spot
becomes anomalously large: $1/\tau_{\bf k}\propto\sqrt
{\varepsilon_{\bf k}}$.  Away from a hot spot, $1/\tau_{\bf
k}\propto\varepsilon_{\bf k}^2$ as in standard Landau's
Fermi-liquid theory.  It is interesting to note that if one
calculates the average of $1/\tau_{\bf k}$ for states of fixed
energy $\varepsilon>T^\star$ along the Fermi line, one obtains
$$\langle{1\over\tau_{\bf k}}\rangle\propto{g^2\over E_F}\:
{\varepsilon\over \omega_D},$$ {\it i.e.}, the average lifetime
of an electron is linear in energy!  It is basically this
feature which has led in previous calculations to the result
$\rho\propto T$ for $T>T^\star$.

\subsection{Resistivity.}
In calculating the temperature dependence of the resistivity due
to spin-fluctuations, we will assume that a simple description
in terms of a Boltzmann equation (BE) captures the essential
physics. Let the stationary solution of the BE be $f_{\bf
k}=f_{\bf k}^\circ-\Phi_{\bf k} (\partial f_{\bf
k}^\circ/\partial\varepsilon),$ where $f_{\bf k}^\circ$ is the
equilibrium Fermi-Dirac distribution function and $\Phi_{\bf k}$
is a function to be determined. The linearized (in $\Phi_{\bf
k}$) collision term of the BE reads
\begin{eqnarray*}
\left({\partial f_{\bf k}\over{\partial t}}
\right)_{\rm coll}&=&{2 g^2\over T}
\sum_{\bf k^\prime}\int_{-\infty}^\infty d\omega\:n(\omega)\:
f_{\bf k^\prime}^\circ(1-f_{\bf k}^\circ)\nonumber\\ &\times&\Im
\chi({\bf k-k^\prime},\omega)\:
(\Phi_{\bf k^\prime}-\Phi_{\bf k})\:
\delta(\varepsilon_{\bf k}-\varepsilon_{\bf k^\prime}-\omega),
\end{eqnarray*}
where $n(\omega)$ is the Bose-Einstein distribution function.
Note that we have assumed that the spin-fluctuations are in
equilibrium.  Following the standard arguments as described by
Ziman \cite{Ziman}, the resistivity $\rho$ can be found as the
minimum of a functional of $\Phi_{\bf k}$:
\begin{equation}
{\rho\over\rho_0}=
\min\left[{\langle\Phi|W|\Phi\rangle\over
\langle\Phi|X\rangle^2}\right],
\label{eq:Zim}
\end{equation}
where we have introduced $\langle\Phi|W|\Phi\rangle=\sum_{\bf
k,k^\prime} W_{\bf k,k^\prime}(\Phi_{\bf k^\prime}-\Phi_{\bf
k})^2$, $\langle\Phi|X\rangle=\sum_{\bf k}\Phi_{\bf k}X_{\bf k}$
and $X_{\bf k}=(-\partial f_{\bf k}^\circ/\partial\varepsilon)
{\bf v}_{\bf k}\cdot {\bf n}$.  $\rho_0=\hbar/e^2$ is the
quantum of resistivity and ${\bf n}$ is a unit vector in the
direction of the applied electric field. Assuming that
$\varepsilon_{\bf k}=\varepsilon_{\bf -k}$ and $\Phi_{\bf k}=
-\Phi_{\bf -k}$, we have $$W_{\bf k,k^\prime}= {2(ga)^2\over T}
f_{\bf k}^\circ(1-f_{\bf k^\prime}^\circ)\: n(\varepsilon_{\bf
k^\prime}-\varepsilon_{\bf k})\:
\Im \chi({\bf k^\prime-k},
\varepsilon_{\bf k^\prime}-\varepsilon_{\bf k}).$$
Similarly as in the discussion of the quasiparticle lifetime, we
write $\int d^2k=\int d\varepsilon\int dk/v$ where the
$k$-integration runs along the Fermi surface, and an analogous
expression for $\int d^2k^\prime$. Defining
$\varepsilon^\prime=\varepsilon+\omega$, we can perform the
$\varepsilon$-integration and obtain
\begin{eqnarray}
{\rho\over\rho_0}&=&\min\left[{{\oint{dk\over v}
\oint{dk^\prime\over v^\prime}F_{\bf k-k^\prime}\:
(\Phi_{\bf k^\prime}-\Phi_{\bf k})^2}\over{
\left(\oint{dk\over v} {\bf v}_{\bf k}\cdot {\bf n}\:
\Phi_{\bf k}\right)^2}}\right]\\
F_{\bf k-k^\prime}&=&{2(ga)^2\over T}\int_0^\infty
d\omega\:\omega\: n(\omega)[n(\omega)+1]
\:\Im \chi({\bf k-k^\prime},\omega).\nonumber
\end{eqnarray}
Note that Eq.2.8 is a generalization of Eq.3.1 from
Ref.\cite{Ueda} to the case of an arbitrary variational function
$\Phi_{\bf k}$.  Using Eq.\ref{eq:fluctuationspectrum} for the
spectrum of spin-fluctuations, we find $F_{\bf k-k^\prime}
=2(ga)^2I(\omega_{\bf k-k^\prime}/T)$, where
$$I(x)=\int_0^\infty{dt\:e^t\over (e^t-1)^2}{t^2\over t^2+x^2}
\approx{\pi^2/3\over{x(x+2\pi/3)}}.$$
The last equality is an interpolation formula which becomes
exact for $x\rightarrow 0$ and $x\rightarrow \infty$.
Summarizing, the sheet resistivity can be written in the
following dimensionless form:
\begin{eqnarray}
&&{\rho\over\rho_0}={\pi^2\over 6}\left({g\over t}\right)^2
\Theta^2\nonumber\\
&&\times\min\left[ {{\oint{dk\over u}\oint{dk^\prime\over
u^\prime} {{(\Phi_{\bf k^\prime}-\Phi_{\bf k})^2}\over
{(\Theta^\star+\alpha\Theta+\psi_{\bf k-k^\prime})
(\Theta^\star+\beta\Theta+\psi_{\bf k-k^\prime})}}}
\over{\left(\oint{dk\over u}{\bf u}_{\bf k}\cdot{\bf n}
\Phi_{\bf k}\right)^2}}\right],
\label{eq:afmresistivity}
\end{eqnarray}
where ${\bf u}_k={\bf v}_{\bf k}/2ta$ is a dimensionless group
velocity, $\Theta=T/\omega_D$, $\Theta^\star=T^\star/\omega_D$, and
$\beta=\alpha+2\pi/3$. The integrations run along the Fermi line. Note
that for $T\ll T^\star/\beta$, the resistivity has a standard
Landau-Fermi-liquid form $\rho\propto T^2$ in agreement with our
results for the quasiparticle lifetime.  In what follows, we will
study the resistivity as given by Eq.\ref{eq:afmresistivity} for
$T^\star/\alpha\ll T\ll
\omega_D$. A standard ansatz for the variational function is
\begin{equation}
\Phi_{\bf k}={\bf u}_{\bf k}\cdot{\bf n}.
\label{eq:standard}
\end{equation}
For such $\Phi_{\bf k}$, there always exists a pair of hot spots
$k,k^\prime$ such that $\Phi_{\bf k^\prime}-\Phi_{\bf k}$ is
finite. In that case, the integral in the nominator of
Eq.\ref{eq:afmresistivity} is dominated by $k,k^\prime$ close to
the hot spots; $\psi_{\bf k-k^\prime}$ can be approximated by a
homogeneous quadratic polynomial in the deviations $\delta
k,\delta k^\prime$ from the hot spots and by scaling, one
obtains $\rho\propto T$ in agreement with Moriya {\it et al.}
\cite{Ueda}.

We have seen that the quasiparticle lifetime is extremely
anisotropic along the Fermi line. It is therefore natural to
assume that the resistivity will be dominated by that part of
the Fermi line where the scattering is weakest and the
contribution from the hot spots will be short-circuited. To
prove this, we take another ansatz for $\Phi_{\bf k}$ and show
that it leads to a lower resistivity. Let us consider
\begin{equation}
\Phi_{\bf k}={{{\bf u}_{\bf k}\cdot{\bf n}}
\over{e^{\beta[\Delta-\varphi]}+1}},
\label{eq:newansatz}
\end{equation}
where $\varphi=\delta k\:a$ is the deviation from a hot spot and
$\Delta$ and $\beta$ are variational parameters. The case
$\beta=0$ corresponds to the standard ansatz, while $\beta\gg 1$
and $\Delta\neq 0$ describe the situation when finite parts of
the Fermi line around the hot spots do not contribute to the
transport.  If $\beta\rightarrow\infty$ and
$\sqrt{T/\omega_D}\ll\Delta\ll 1$, we obtain at low temperatures
$\rho\propto T^2$ even for $T^\star=0$, since
$\langle\Phi|W|\Phi\rangle$ becomes temperature-independent.
With increasing temperature, one is forced to choose larger
$\Delta$ in order to exclude the hot-spot regions. This leads,
however, to a decrease of $\langle\Phi|X\rangle$ and finally at
high enough temperatures, the solution Eq.\ref{eq:newansatz}
with a large $\beta$ becomes unfavourable compared to the
standard ansatz Eq.\ref{eq:standard}. This happens if
$T/\omega_D>c$, where $c$ is a numerical factor which depends on
the details of the geometry of the Fermi line and of the hot
spots. We were unable to make a reliable estimate of $c$
analytically and therefore we calculated $\rho$ as a function of
$T$ numerically.

In Fig.2, we show the results of a numerical calculation of the
resistivity according to Eq.\ref{eq:afmresistivity} using both
the standard and improved ansatz for $\Phi_{\bf k}$.  For the
spin-fluctuation spectrum we take $T^\star=0,\alpha=2.0$, and
$\omega_D=1760 K$. The density of electrons is $n=0.75$ and the
coupling constant $g=0.64 eV$. Note that with the standard
ansatz Eq.\ref{eq:standard}, we obtain for this critical system
$\rho\propto T$ down to $T=0$ in agreement with previous studies
\cite{Ueda} (see also \cite{Hartmut}). With the improved ansatz
Eq.\ref{eq:newansatz}, the resistivity is lower for all studied
temperatures and it is proportional to $T^2$ up to $T\approx 70
K$. For $T>70 K$, $\rho$ is a linear function of $T$ with a
similar slope as for the standard ansatz.  However, the
extrapolation of the linear part down to $T=0$ is negative.

Finally, let us consider the spin-fluctuation spectrum with
$T^\star=110 K$, $\alpha=0.55$, and $\omega_D=1760 K$.  We take
again $n=0.75$ and $g=0.64 eV$.  These are the same parameters
as those used in Ref.\cite{MP} (see their Eq.37 and note that
$\omega_D=2\omega_{SF}(\xi/a)^2$; $\omega_{SF}$ and $\xi$ are
the parameters for the spin-fluctuation spectrum used in
Ref.\cite{MP}).  The results of our numerical calculation are
shown in Fig.3. The standard ansatz Eq.\ref{eq:standard} yields
a resistivity-vs.-temperature curve qualitatively similar to
that obtained by Monthoux and Pines \cite{MP}, but our
resistivity is approximately three times larger \cite{Note}.
$\rho$ is a linear function of $T$ down to $T\approx 100K$. Our
improved solution Eq.\ref{eq:newansatz} yields smaller values of
resistivity: for instance, $\rho(T_c)$ calculated using our
ansatz is only $\approx 0.6$ of the value obtained with the
standard ansatz.  More importantly, the shape of the $\rho(T)$
curve changes: it is linear only above $T\approx 180 K$.  We
believe that the latter feature will hold true also in a more
sophisticated calculation than in our Boltzmann-equation
approach.  In order to proceed further in this direction it will
be necessary to find a translation of the variational principle
used here to the Green's-function formulation of transport
problems.

\subsection{Influence of impurity scattering
on the resistivity.} In the presence of impurities, one can expect
that the anisotropy of the quasiparticle lifetime will be suppressed.
Thus, a question arises what is the actual temperature dependence of
the resistivity in such a case \cite{Kazuo}. We will address this
question by assuming that the impurity scattering can be described by
the Boltzmann equation (thus disregarding all fully quantum-mechanical
effects like weak localization, etc.) In that case, the resistivity
can be described by Eq.2.8 where $F_{\bf k-k^\prime}$ acquires an
additional contribution $F_{\bf k-k^\prime}^{\rm imp}$ from impurity
scattering.  In the Born approximation, $F_{\bf k-k^\prime}^{\rm
imp}=\pi a^2 |H^\prime_{\bf k,k^\prime}|^2$, where $H^\prime$
describes the interaction of an electron with impurities.  Since we
are not interested here in a microscopic calculation of the
resistivity due to impurities $\rho_{\rm imp}$, we take
$|H^\prime_{\bf k,k^\prime}|=V$ where $V$ is a free parameter to be
chosen so as to give a realistic $\rho_{\rm imp}$.
Under these assumptions, we have
\begin{equation}
{\rho_{\rm imp}\over\rho_0}={\pi\over 2}\left({V\over t}\right)^2
{{\langle\langle\Phi_{\bf k}^2\rangle\rangle}\over
{\langle\langle\Phi_{\bf k}{\bf u_k}\cdot{\bf n}\rangle\rangle^2}},
\end{equation}
where $\langle\langle A\rangle\rangle=\oint{dk\over v}A_k/
\oint{dk\over v}$ is an average of $A$ along the Fermi surface.
It is easy to see that $\rho_{\rm imp}$ is minimized by the standard
ansatz Eq.\ref{eq:standard}. Since the resistivity due to impurities
is finite down to $T=0$ while the contribution of spin-fluctuations
vanishes in that limit, it is clear that the standard ansatz will be
favourable for $T\rightarrow 0$. At higher temperatures, however, the
decrease of the spin-fluctuation contribution to the resistivity for
$\Phi_{\bf k}$ given by Eq.\ref{eq:newansatz} may outweigh the
increase of the contribution due to impurities. In order to test this
possibility, we performed a calculation of the resistivity with the
same parameters as those used in Fig.3; we assumed a residual
resistivity $\rho_{\rm imp}(T=0)=0.25\rho_0$. The result of this
calculation is shown in Fig.4. It is seen that the presence of
impurity scattering decreases the difference between the resistivity
as calculated by the standard ansatz Eq.\ref{eq:standard} and our
variational function Eq.\ref{eq:newansatz}. However, even for the
relatively large impurity scattering we have chosen, the resistivity
is still not a linear function of temperature for $T>100K$.

Summarizing the results of the present Section we can say that
although the quasiparticle lifetime is anomalous around special
points on the Fermi line (hot spots), the resistivity is
proportional to $T^2$ at low enough temperatures. The energy
scale where a crossover to $\rho\propto T$ occurs is determined
not only by the parameter $T^\star$ as found in previous
studies, but also by some fraction of $\omega_D$.

\section{Models with van Hove singularities on the Fermi line.}
In the literature there appeared a number of attempts to explain
the anomalies of the normal state of the HTc's in the framework
of a weak-coupling theory under the assumption of a special
single-particle dispersion. The most promising among these are
the models which assume that at the optimal doping, there is a
van Hove singularity on the Fermi line.  Such singularities
always exist in a periodic energy band for topological reasons
(see, {\it e.g.}, Ref.\cite{JM}). The special feature of the
HTc's according to these theories simply is that a crossing of
the Fermi line with a van Hove singularity can occur. In the
present Section, we will calculate the quasiparticle lifetime
and the resistivity in the van Hove scenario.  We consider
electrons with the Hamiltonian
\begin{equation}
H=\sum_{{\bf k},\sigma}\varepsilon_{\bf k} c^\dagger_{{\bf
k},\sigma} c_{{\bf k},\sigma}+g\sum_i n_{i,\uparrow}
n_{i,\downarrow},
\end{equation}
where $\varepsilon_{\bf k}$ is the single-particle dispersion
Eq.\ref{eq:MPdispersion} and $g$ is a weak screened interaction
among the electrons. It is assumed that $t>2t^\prime>0$. In that
case, there exist two saddle-points $(\pi/a,0)$ and $(0,\pi/a)$
which lead to a van Hove singularity at energy $-4t^\prime$.
The Fermi line for a filling when it goes through a van Hove
singularity is shown in Fig.5. The single-particle dispersion
around the saddle points becomes anomalous: e.g., in the
neighborhood of $(\pi/a,0)$, we have
\begin{equation}
\varepsilon_{\bf k}\approx
k_y^2/2m_y-(k_x-\pi/a)^2/2m_x,
\label{eq:SPspectrum}
\end{equation}
where $1/m_x=(t-2t^\prime)a^2$ and $1/m_y=(t+2t^\prime)a^2$.

\subsection{Quasiparticle lifetime.}
Before considering the temperature dependence of the resistivity in
the van Hove scenario, let us first calculate the quasiparticle
lifetime. We rederive here the results of Gopalan {\it et al.}
\cite{Sudha} in a simpler way which will enable us to calculate the
resistivity in the next subsection.  The quasiparticle lifetime is
given by Eq.\ref{eq:goldenrule1} where $\Im\chi({\bf
q},\omega)=\pi\sum_{\bf k} f_{\bf k}(1-f_{\bf
k+q})\delta(\varepsilon_{\bf k+q}-\varepsilon_{\bf k}-\omega)$ is the
imaginary part of the bare susceptibility.  In what follows, we study
the lifetime of an electron ${\bf k}$ which is scattered to ${\bf
k^\prime}$ by exciting a particle-hole pair ${\bf K}\rightarrow{\bf
K^\prime}$.

Let us first consider the case when all involved states ${\bf
k,k^\prime,K}$, and ${\bf K^\prime}$ are in the neighborhood of the
saddle points. There are two types of such scatterings: intra- and
inter-saddlepoint scatterings (with or without umklapp), respectively.
If both ${\bf K}$ and ${\bf K^\prime}$ lie close to the
same saddle point, we can use the expression for the susceptibility
found by Gopalan {\it et al.} \cite{Sudha} (for $T=0$):
\begin{equation}
\Im\chi({\bf q},\omega)\propto
\min\left[1,{\omega\over{|\varepsilon_{\bf q}|}}\right],
\label{eq:SPsusceptibility}
\end{equation}
where $\varepsilon_{\bf q}$ is given by Eq.\ref{eq:SPspectrum}.
To simplify the analysis of the susceptibility in the case when
${\bf K}$ and ${\bf K^\prime}$ are close to different saddle points,
let us assume for the moment that the single-particle spectrum is
\begin{equation}
\varepsilon_{\bf k}={1\over 2m}\left(|k_x|-{G\over 2}\right)k_y.
\label{eq:umklappspectrum}
\end{equation}
The Fermi line for electrons with the dispersion
Eq.\ref{eq:umklappspectrum} is shown in Fig.6. The spectrum consists
of two saddle points whose distance is ${\bf G}=(G,0)$, where $2{\bf
G}$ is assumed to be a vector of the inverse lattice.  The
susceptibility at a wavevector ${\bf G-q}$ where ${\bf q}$ is small
then is $$
\Im\chi({\bf G-q},\omega)\propto|
\ln|{2\omega\over{|\varepsilon_{\bf q}|}}-{\rm sgn}
(\varepsilon_{\bf q})||, $$ where $\varepsilon_{\bf q}=q_xq_y/2m$.
The lifetime of the electron in the state ${\bf k}$ now reads
$${1\over \tau_{\bf k}}=2g^2\sum_{\bf k^\prime}^\prime\Im\chi({\bf
k-k^\prime},\varepsilon_{\bf k}-\varepsilon_{\bf k^\prime}),$$ where
the prime on the sum means a restriction to those states ${\bf
k^\prime}$ which satisfy $0<\varepsilon_{\bf
k^\prime}<\varepsilon_{\bf k}$. One finds from here that at
zero temperature $1/\tau_{\bf k}\propto\varepsilon_{\bf k}$ in
agreement with the result of Gopalan {\it et al.}\cite{Sudha}.

Let us calculate now the lifetime of an electron ${\bf k}$ away
from the saddle points when two of the states ${\bf
k^\prime,K}$, and ${\bf K^\prime}$ are close to a saddle point.
We make use of Eq.\ref{eq:goldenrule2}.  Let us consider first
the contribution of ${\bf k^\prime}$ close to ${\bf k}$
(forward-scattering channel) for which the expression
Eq.\ref{eq:SPsusceptibility} is valid. Let $|{\bf
k-k^\prime}|=q$.  Since none of the asymptotes of the hyperbolas
Eq.\ref{eq:SPspectrum} is parallel to the Fermi line (away from
the saddle points), we have $${1\over\tau_{\bf
k}}\propto\int_0^{\varepsilon_{\bf k}} d\omega\int_0^\Lambda
dq\min\left[1,|M\omega/q^2|\right],$$ where $M$ is a constant
and $\Lambda$ a cut-off in momentum space.  Taking the integral,
we have $1/\tau_{\bf k}\propto
\varepsilon_{\bf k}^{3/2}$ in agreement with
Ref.\cite{Sudha}.  Let us consider now ${\bf k^\prime}$ close to a
saddle point.  If we require that one of the points ${\bf K}$ and
${\bf K^\prime}$ is close to a saddle point, we find that either ${\bf
K}\approx{\bf k^\prime}$ and ${\bf K^\prime}\approx{\bf k}$ (exchange
channel) or ${\bf K}\approx{-\bf k}$ and ${\bf K^\prime}\approx{-\bf
k^\prime}$ (Cooper channel).  The contribution to the lifetime of the
exchange channel is analogous to that of the forward-scattering
channel.  In order to calculate the contribution of the Cooper channel
we need to calculate the susceptibility $\Im\chi({\bf
K^\prime-K},\omega)$ where ${\bf K}$ and ${\bf K^\prime}$ are momenta
away and close to a saddle point, respectively. Let ${\bf P}$ and
${\bf Q}$ be points on the Fermi line close to ${\bf K}$ and ${\bf
K^\prime}$, respectively such that ${\bf Q-P}={\bf K^\prime-K}$. Let
the spectrum in the vicinity of the saddle point and of ${\bf P}$ be
$\varepsilon_{\bf k}=k_xk_y/2m$ and $\varepsilon_{\bf k}={\bf
v}\cdot({\bf k-P})$, respectively and let ${\bf Q}=(Q,0)$. Then the
susceptibility is
\begin{equation}
\Im\chi({\bf K^\prime-K},\omega)\approx
{1\over(2\pi)^2v\cos\phi}
\left(\sqrt{Q^2+{8m\omega\over\tan\phi}}-Q\right),
\label{eq:newsusc}
\end{equation}
where $\phi$ is the angle between the tangents to the Fermi line in
the points ${\bf P}$ and ${\bf Q}$. Now we can calculate the lifetime
in the Cooper channel according to Eq.\ref{eq:goldenrule1} by first
integrating over $\omega$ and introducing hyperbolic coordinates
${\bf k^\prime}=(\varepsilon^\prime,
\phi^\prime)$
such that $k^\prime_x\propto\sqrt{\varepsilon^\prime/\tan\phi^\prime}$
and $k^\prime_y\propto \sqrt{\varepsilon^\prime\tan\phi^\prime}$.
$Q$ is dominated by the position of ${\bf k^\prime}$
and we find $Q\propto\sqrt{\varepsilon^\prime/\tan\phi^\prime}
|\tan\phi^\prime-\tan\phi|$. The resulting integral can be
performed by scaling and we find $1/\tau_{\bf k}\propto
\varepsilon_{\bf k}^{3/2}$.

Finally, in case when ${\bf k}$ is away from saddle points and
at most one of the points ${\bf k^\prime,K}$, and ${\bf
K^\prime}$ is close to a saddle point we obtain a lifetime
analogous to the result for an isotropic spectrum
\cite{Wilkins}. As an example, let us consider the case
when it is the momentum ${\bf K^\prime}$ which is close to a
saddle point. We calculate the lifetime according to
Eq.\ref{eq:goldenrule2} where the susceptibility is given by
Eq.\ref{eq:newsusc}. Let ${\bf k^\prime_0}$ be that value of
${\bf k^\prime}$ for which $Q=0$. For a general ${\bf k^\prime}$
we have $Q=\alpha q$ where $q=|{\bf k^\prime-k^\prime_0}|$ and
$\alpha$ is a constant.  The lifetime is $${1\over \tau_{\bf
k}}\propto
\int_0^{\varepsilon_{\bf k}}d\omega\int_0^\Lambda dq
\left(\sqrt{q^2+{8m\omega\over{\alpha^2\tan\phi}}}
-q\right),$$ where $\Lambda$ is a cut-off in momentum space. The
integration is straightforward and we find $1/\tau_{\bf
k}\propto\varepsilon_{\bf k}^2\ln(1/\varepsilon_{\bf k})$.  Thus
the contribution to $1/\tau_{\bf k}$ of the processes with one
of the scattering states close to a saddle point is subleading
compared to the processes in the forward-scattering, exchange,
and Cooper channels.

Summarizing, we have found that due to the presence of van Hove
singularities on the Fermi line, the scattering rate is
anomalously enhanced compared to the isotropic case; for
electrons close to a saddle point $1/\tau\propto\varepsilon$,
whereas for the remaining electrons
$1/\tau\propto\varepsilon^{3/2}$.

\subsection{Resistivity.}
Let us calculate the resistivity in the van Hove scenario.  We
will work again in the quasiclassical formalism of the Boltzmann
equation. Analogously to the discussion in Section 2, the
resistivity can be found \cite{Ziman} as a minimum of the
functional Eq.\ref{eq:Zim} where
\begin{eqnarray}
&&\langle\Phi|W|\Phi\rangle={\pi g^2\over 2T}\sum_{\bf G}
\sum_{\bf k,k^\prime,K,K^\prime}
f_{\bf k}f_{\bf K}(1-f_{\bf k^\prime})(1-f_{\bf K^\prime})
\nonumber\\
&&\times\left(\Phi_{\bf k}+\Phi_{\bf K}-\Phi_{\bf k^\prime}
-\Phi_{\bf K^\prime}\right)^2\nonumber\\
&&\times\delta(\varepsilon_{\bf k}+\varepsilon_{\bf K}
-\varepsilon_{\bf k^\prime}-\varepsilon_{\bf K^\prime})
\delta({\bf k+K-k^\prime-K^\prime-G})
\label{eq:Zimresistivity}
\end{eqnarray}
and ${\bf G}$ is a reciprocal-lattice vector. In what follows we
consider the standard ansatz $\Phi_{\bf k}={\bf v}_{\bf
k}\cdot{\bf n}$ and we neglect the weak temperature dependence
of $\langle\Phi|X\rangle$. Then we can write Eq.\ref{eq:Zim} in
the following form: $$\rho\propto{1\over T}\sum_{\bf k}{f_{\bf
k}\over\tau_{\bf k}^{TR}}\sim\oint{dk\over v}{1\over
\tau^{TR}(k,T)},$$
{\it i.e.}, the resistivity can be found as an average over the
Fermi line of the transport scattering rate at energy $\sim T$.
For the transport lifetime we find an expression similar to that
for the quasiparticle lifetime:
\begin{eqnarray*}
{1\over \tau_{\bf k}^{TR}}&\propto&\sum_{\bf k^\prime}
\int_0^{\varepsilon_{\bf k}}d\omega\:
\Im\chi^{TR}({\bf k-k^\prime},\omega;
\Phi_{\bf k}-\Phi_{\bf k^\prime})\\ &&\times\delta
(\varepsilon_{\bf k}-\varepsilon_{\bf k^\prime}-\omega),\\
\Im \chi^{TR}({\bf q},\omega;u)&=&\pi\sum_{\bf k}f_{\bf k}
(1-f_{\bf k+q})(\Phi_{\bf k+q}-\Phi_{\bf k}-u)^2\\&&\times
\delta(\varepsilon_{\bf k+q}-\varepsilon_{\bf k}-\omega),
\end{eqnarray*}
where we have defined the `transport susceptibility' $\Im
\chi^{TR}({\bf q},\omega;u)$.

Let us study first the transport lifetime for the processes when all
electron states involved in the scattering are close to the saddle
points.  Note that assuming intra-saddle-point scatterings and a
dispersion $\varepsilon_{\bf q}=q_x q_y/2m$, the conservation of
momentum implies $\Phi_{\bf k}+\Phi_{\bf K}-\Phi_{\bf k^\prime}
-\Phi_{\bf K^\prime}=0$ similarly as in the case of an isotropic
dispersion and the resistivity vanishes \cite{Newns}.  Thus we have to
take into account inter-saddle-point scatterings.  Unfortunately, the
transport lifetime for the model dispersion
Eq.\ref{eq:umklappspectrum} is different from the actual result for
the spectrum Eq.\ref{eq:MPdispersion}, since the asymptotes of the
hyperbolas of the two saddle points in the latter spectrum are not
parallel to each other and we have to calculate more carefully.  Let
us assume without loss of generality that ${\bf k^\prime}$ lies in the
vicinity of the point $(\pi/a,0)$ where the dispersion is described by
Eq.\ref{eq:SPspectrum}. The energy conservation together with the
Pauli principle require $0<\varepsilon_{\bf k^\prime}<\varepsilon_{\bf
k}$ and the allowed ${\bf k^\prime}$-points lie between two branches
of a hyperbola centered at $(\pi/a,0)$.  The dominant contribution to
$1/\tau_{\bf k}^{TR}$ comes from the ${\bf k^\prime}$-points in the
tails of the hyperbolas, where the transport susceptibility
$\Im\chi^{TR}({\bf k-k^\prime},\varepsilon_{\bf k}-\varepsilon_{\bf
k^\prime})\approx \varepsilon_{\bf k}$. Thus the transport lifetime
reads $${1\over\tau_{\bf k}^{TR}}\propto\varepsilon_{\bf
k}\int_0^\Lambda dk^\prime_x\int_{\alpha k^\prime_x}^{\sqrt{(\alpha
k^\prime_x)^2+2m_y\varepsilon_{\bf k}}}dk^\prime_y,$$ where
$\alpha=\sqrt{m_y/m_x}$, $\Lambda$ is a cut-off in momentum space
(typically some fraction of $\pi/a$) and we have shifted the position
of the saddle point to $(0,0)$. Taking the integral we find
$1/\tau_{\bf k}^{TR}\propto \varepsilon_{\bf
k}^2\ln(\Lambda^2/2m_x\varepsilon_{\bf k})$.

Now we consider processes when two states lie close to a saddle point.
As discussed in the previous subsection there are three types of such
processes, namely scatterings in the forward, exchange, and Cooper
channel. In the forward-scattering channel, the exchanged momentum
${\bf q}$ is small and we can estimate $(\Phi_{\bf k+q}-\Phi_{\bf
k}-u)^2\sim {\bf q}^2$.  Repeating the analysis of the previous
subsection we find $${1\over\tau^{TR}_{\bf
k}}\propto\int_0^{\varepsilon_{\bf k}} d\omega\int_0^\Lambda dq
q^2\min\left[1,|M\omega/q^2|\right]
\propto\varepsilon_{\bf k}^2.$$ Scattering in the exchange
channel leads to a similar result.  In the Cooper channel the
relevant transport susceptibility has an additional factor
$(\Phi_{\bf k}+\Phi_{\bf K} -\Phi_{\bf k^\prime}-\Phi_{\bf
K^\prime})^2\propto\varepsilon_{\bf k}$ compared to
Eq.\ref{eq:newsusc} and hence $1/\tau^{TR}_{\bf k}\propto
\varepsilon_{\bf k}^{5/2}$.

Finally, if only one of the states ${\bf k,k^\prime,K}$, and
${\bf K^\prime}$ is close to a saddle point there is in general
no additional small factor distinguishing $1/\tau$ from
$1/\tau^{TR}$ and thus the resistivity is $\rho\propto
T^2\ln(1/T)$. Note that although processes of this type give
only a subdominant contribution to the quasiparticle lifetime,
they provide a leading contribution to relaxation of momentum.

Summarizing, in Section 2 we calculated the quasiparticle
lifetime and the resistivity in the van Hove scenario. We found
that although the quasiparticle lifetime is anomalously short
the resistivity exhibits the standard temperature dependence
with a logarithmic correction $\rho\propto T^2\ln(1/T)$ for
$T\ll E_F$.

\section{Conclusions.}
In this paper we have analyzed the resistivity as a function of
temperature for two two-dimensional models with hot spots:
nearly antiferromagnetic Fermi liquids and a model with van Hove
singularities on the Fermi line. To simplify the treatment, we
decided to formulate the transport problem on the level of a
Boltzmann equation.

In the case of nearly antiferromagnetic Fermi liquids, we have shown
that the standard treatment which does not take into account the
anisotropy of the electron lifetime along the Fermi line leads to
$\rho\propto T$ for $T>T^\star$.  However, we constructed better
variational solutions of the BE which exclude highly resistive points
on the Fermi line and yield $\rho\propto T^2$ even above $T^\star$.
We have found a new energy scale for the crossover to the $\rho\propto
T$ behavior at higher temperatures. This energy scale does not vanish
even if the spectrum of the spin-fluctuations is critical.  The
presence of disorder was shown to decrease the difference between the
standard solution and our ansatz; however, even for relatively strong
disorder, the resistivity is not a linear function of temperature
above 100K, if we use the parameters proposed by Monthoux and Pines
\cite{MP}.

More generally, our analysis suggests that if the electrons couple to
a bosonic excitation which is soft at some {\it finite} wavevector
${\bf Q}$, the quasiparticle lifetime will be very anisotropic and
there will be hot spots on the Fermi line where the scattering may
become anomalous. However, the resistivity will be dominated by the
lifetime in the generic points of the Fermi line away from the hot
spots. Thus, only anomalous scattering in a generic point on the Fermi
line implies anomalous resistivity.  This can be achieved, {\it e.g.},
by coupling the electrons to a mode which is soft at long wavelengths.
For example, consider electrons with the spectrum $\varepsilon_{\bf
k}={\bf k}^2/2m$ coupled to bosons described by the propagator
$$\chi({\bf q},\omega)\propto {q^\beta\over{\omega_{\bf q}-i\omega}}
\hspace{0.3cm}{\rm or}\hspace{0.3cm}
\chi({\bf q},\omega)\propto {q^\beta\omega_{\bf q}\over{
\omega^2_{\bf q}-\omega^2}},$$ where $\omega_{\bf
q}=T^\star+Aq^\alpha$; we take $\alpha\geq 1$, $\beta\geq 0$.  A
golden-rule calculation of the quasiparticle lifetime and resistivity
for $T>T^\star$ gives $1/\tau\propto T^{(D+\beta-1)/\alpha}$ and
$\rho\propto T^{(D+\beta+1)/\alpha}$, respectively.  $D$ is the
spatial dimension (we assume that $D$ is the same for both electrons
and bosons).  {\it E.g.}, for scattering of spinons on the gauge field
we have $D=2$, $\alpha=3$, $\beta=1$ and we obtain $1/\tau\propto
T^{2/3}$ and $\rho\propto T^{4/3}$ in agreement with Ref.\cite{NL}.
For $T\ll T^\star$, we have $1/\tau\propto T^2$. Thus the concept of
quasiparticles is valid and our use of a semiclassical approximation
is supported.  It is interesting to note that for $T^\star=0$,
$\rho\propto T$ would require $1/\tau\propto T^\nu$ with $\nu<1$;
coupling the electrons to a bosonic excitation and requiring that
Landau's Fermi-liquid theory is applicable would lead to $\rho\propto
T^\mu$ where $\mu>(2+\alpha)/\alpha$.

In the second part of this paper, we considered a model with van
Hove singularities close to the Fermi line and with weak
screened electron-electron interactions.  If the van Hove
singularity is located at energy $E_F\pm T^\star$, then the
anomalous quasiparticle lifetime reported in Section 3 is valid
for $\varepsilon\gg T^\star$. At energies smaller than $T^\star$
we obtain the standard results for interacting electrons in two
dimensions \cite{Wilkins} and the concept of quasiparticles
should be applicable. Thus the existence of a nonvanishing
$T^\star$ provides, similarly as in the case of nearly
antiferromagnetic Fermi liquids, support for our use of the
semiclassical approach.  At $\varepsilon\gg T^\star$,
small-angle scatterings or Cooper-channel processes in which
electrons close to saddle points take part are responsible for
the anomalous behavior of the quasiparticle lifetime. However,
using the standard ansatz $\Phi_{\bf k}={\bf v}_{\bf k}\cdot{\bf
n}$ for the variational solution of the Boltzmann equation their
contribution to the transport lifetime becomes regularized by
the appearance of an additional small factor $(\Phi_{\bf
k}+\Phi_{\bf K}-\Phi_{\bf k^\prime}-\Phi_{\bf K^\prime})^2$ in
Eq.\ref {eq:Zimresistivity} and we obtain finally $\rho\propto
T^2\ln(1/T)$. Such a reduction of the transport scattering rate
compared to the quasiparticle scattering is in fact quite common
as can be seen, {\it e.g.}, from our results for electrons
interacting with a bosonic mode. Another example is a
one-dimensional system away from half filling: the quasiparticle
lifetime behaves as $1/\tau\propto T$, while the resistivity is
exponentially small \cite{Giamarchi}. A similar tendency holds
for electrons with the usual isotropic dispersion in two
dimensions: Hodges {\it et al.} \cite{Wilkins} found that
$1/\tau\propto T^2\ln(1/T)$, whereas it can be shown
\cite{Fujimoto} that the resistivity $\rho\propto T^2$.

\acknowledgements
We would like to thank G. Blatter, H. Monien, A. Ruckenstein,
H. Tsunetsugu, and K. Ueda for interesting discussions.


\begin{figure}
Fig.1. Solid line: the Fermi line for the spectrum
Eq.\ref{eq:MPdispersion}.  We take $t^\prime=0.45 t$ and $n=0.75$.
The hot spots satisfying the condition Eq.\ref{eq:hotspot} are located
in the cross-section points of the Fermi line with the dashed square.
\end{figure}

\begin{figure}
Fig.2. Resistivity due to spin-fluctuations as a function of
temperature for the spin-fluctuation parameters $T^\star=0$,
$\alpha=2.0$, $\omega_D=1760 K$, the electron parameters $t=0.25 eV$,
$t^\prime=0.45 t$, $n=0.75$, and the coupling constant $g=0.64 eV$.
Solid line: calculation with the improved ansatz
Eq.\ref{eq:newansatz}.  Dashed line: calculation with the standard
ansatz Eq.\ref{eq:standard}.
\end{figure}

\begin{figure}
Fig.3. Resistivity due to spin-fluctuations as a function of
temperature for the spin-fluctuation parameters $T^\star=110 K$,
$\alpha=0.55$, $\omega_D=1760 K$, the electron parameters $t=0.25 eV$,
$t^\prime=0.45 t$, $n=0.75$, and the coupling constant $g=0.64 eV$.
Solid line: calculation with the improved ansatz
Eq.\ref{eq:newansatz}.  Dashed line: calculation with the standard
ansatz Eq.\ref{eq:standard}.
\end{figure}

\begin{figure}
Fig.4. Resistivity due to spin-fluctuations as a function of
temperature for the same parameters as in Fig.3.  Additional impurity
scattering $\rho_{\rm imp}(T=0)=0.25\rho_0$ was assumed.  Solid line:
calculation with the improved ansatz Eq.\ref{eq:newansatz}.  Dashed
line: calculation with the standard ansatz Eq.\ref{eq:standard}.
\end{figure}

\begin{figure}
Fig.5. Fermi line for the spectrum Eq.\ref{eq:MPdispersion}.  We
take $t^\prime=0.28 t$ and $n=0.75$. The van Hove singularities
are located in the points $(\pi/a,0)$ and $(0,\pi/a)$.
\end{figure}

\begin{figure}
Fig.6. Fermi line for the spectrum Eq.\ref{eq:umklappspectrum}.
The shaded regions show the occupied states.
\end{figure}

\end{document}